\documentclass[11pt,a4paper]{article}
\usepackage{graphicx}
\usepackage{jcappub,natbib}
\usepackage{braket}

\title{Parity-violating and anisotropic correlations in pseudoscalar inflation}
\author[a,b]{Nicola Bartolo,}
\author[a,b,c]{Sabino Matarrese,}
\author[d]{Marco Peloso}
\author[a,b]{and Maresuke Shiraishi}

\affiliation[a]{Dipartimento di Fisica e Astronomia ``G. Galilei'', \\ 
Universit\`a degli Studi di Padova, via Marzolo 8, I-35131, Padova, Italy}
\affiliation[b]{INFN, Sezione di Padova, \\ 
via Marzolo 8, I-35131, Padova, Italy}
\affiliation[c]{Gran Sasso Science Institute, INFN, \\ 
viale F. Crispi 7, I-67100, L'Aquila, Italy}
\affiliation[d]{School of Physics and Astronomy, \\ 
University of Minnesota, Minneapolis 55455, USA}

\abstract{%
A pseudo-scalar inflaton field can have interesting phenomenological signatures associated with parity violation. The existing analyses of these signatures typically assume statistical isotropy. In the present work we instead investigate the possibility that a pseudo-scalar inflaton is coupled to a vector field carrying a small but non-negligible vacuum expectation value (vev) coherent over our Hubble patch.  We show that, in such case, correlators involving the primordial curvature perturbations and gravitational waves violate both statistical isotropy and parity symmetry. We compute the Cosmic Microwave Background (CMB) temperature anisotropies (T) and polarization (E/B) generated by these primordial modes. The CMB two-point correlation functions present distinct signals of broken rotational and parity invariance. Specifically, we find non-vanishing TT, TE, EE and BB correlators between $\ell_1$ and $\ell_2 = \ell_1 \pm 1$ multipoles, and non-vanishing  TB and EB correlators between $\ell_1$ and $\ell_2 = \ell_1 \pm 2$ multipoles. Such signatures are specific of the models under consideration and they cannot be generated if one of parity and isotropy is preserved. As a specific example we consider the simple case in which the vector field has just an ``electric'' background component decaying in the standard way as $a^{-2}$. In this case a strong scale-dependent quadrupolar modulation of the primordial power spectra is generated and we find that almost noiseless data of the large-scale temperature and E-mode polarization anisotropies (like, e.g., the ones provided by WMAP or {\it Planck}) should be able to constrain the quadrupolar amplitude coefficients $g_{2M}$ of the primordial scalar power spectrum (normalized at the pivot scale comparable to the present horizon size $k_0^{-1} = 14~{\rm Gpc}$) down to $g_{2M} = 30$ ($68 \%$CL).}

\begin{document}

\begin{flushright}
{\small UMN-TH-3409/14}
\end{flushright}

\maketitle
\flushbottom

\section{Introduction}

Pseudo-scalar fields with axial symmetry are very common in particle physics, as they often arise from the breaking of global symmetries.  In the cosmological context, UV-complete inflationary models where such a pseudoscalar is identified as the inflaton or a particle affecting inflationary dynamics have been widely discussed e.g., \cite{Freese:1990rb,Adams:1992bn,Lue:1998mq,Kim:2004rp,Dimopoulos:2005ac,McAllister:2008hb,Kaloper:2008fb,Kaloper:2011jz, Pajer:2013fsa}. Recently, they also started drawing attention in models achieving cosmological magnetogenesis \cite{Garretson:1992vt, Son:1998my, Field:1998hi, Vachaspati:2001nb,Sigl:2002kt, Durrer:2010mq,  Caprini:2014mja, Cheng:2014kga}.

Pseudoscalar inflation (involving the $f(\phi) F\tilde{F}$ term) imprints interesting signatures in cosmological phenomena \cite{Lue:1998mq, Barnaby:2010vf, Sorbo:2011rz, Barnaby:2011qe, Barnaby:2011vw, Maleknejad:2011jw, Maleknejad:2011sq, Barnaby:2012xt, Maleknejad:2012fw, Cook:2013xea, Shiraishi:2013kxa, Maleknejad:2014wsa, Mukohyama:2014gba, Ferreira:2014zia, Ozsoy:2014sba}. In these scenarios inflationary gravitational waves can break parity being sourced by the U(1) gauge field in which one of the two polarization states gets enhanced (due to the coupling with the pseudoscalar field). They are directly reflected in observed temperature (T) and polarization (E/B) anisotropies of the Cosmic Microwave Background (CMB). Especially, nonzero cross correlations between temperature/E-mode and B-mode anisotropies (TB/EB) are powerful evidences of parity-violating gravitational waves \cite{Lue:1998mq,Sorbo:2011rz, Barnaby:2011vw, Cook:2013xea, Shiraishi:2013kxa, Maleknejad:2014wsa}.

Previous studies on CMB phenomenology in pseudoscalar inflation models are based on an assumption of statistical isotropy of the Universe.\footnote{See refs.~\cite{Dimopoulos:2012av, Maleknejad:2013npa} for other works that discuss violation of both parity and statistical anisotropy in inflationary perturbations.} In such cases, CMB power spectra are restricted by a diagonal condition in $\ell$ space ($\ell_1 = \ell_2$) due to rotational invariance \cite{Sorbo:2011rz, Barnaby:2011vw, Cook:2013xea, Shiraishi:2013kxa}. This work focuses on pseudoscalar inflation models endowed with a small statistical anisotropy, and as a consequence we find that specific nonzero off-diagonal correlations appear (see table~\ref{tab:correspond} for a summary). 

If one or more vector fields are dynamically relevant during inflation (irrespectively of whether the inflaton is a scalar or a pseudoscalar), they generally acquire a non-vanishing  vacuum expectation value (vev) coherent on super-horizon scales. This leads to  anisotropic expansion and can imprint a  directional dependence on the CMB polyspectra, including non-vanishing off-diagonal signals in the  CMB covariance matrix $\langle a_{\ell_1 m_1} a_{\ell_2 m_2}^{*} \rangle$, e.g., \cite{Ackerman:2007nb, Watanabe:2010bu, Shiraishi:2011ph, Bartolo:2011ee}. For example, in the well-known $f(\phi) F^2$ model, a quadrupolar modulation arises in correlations of primordial perturbations \cite{Himmetoglu:2009mk, Watanabe:2010fh, Gumrukcuoglu:2010yc, Dulaney:2010sq, Bartolo:2012sd, Shiraishi:2013vja, Shiraishi:2013oqa}, and the amplitude parameters of these modulations (conventionally denoted by $g_*$ for the power spectrum 
\cite{Ackerman:2007nb} and $c_2$ for the bispectrum \cite{Shiraishi:2013vja})  have been constrained by CMB observations to be consistent with 0 \cite{Ade:2013ydc,Kim:2013gka}. In this model, rotational invariance is broken, but parity symmetry is conserved; hence nonzero TT, TE, EE and BB (TB and EB) are allowed for the configurations $\ell_1 = \ell_2 \pm 2, \ell_2$ ($\ell_1 = \ell_2 \pm 1$). 

A  goal of this work is to explore primordial scalar (curvature perturbation) and tensor (gravitational waves) correlations and their signatures on CMB power spectra in the cases in which a pseudo-scalar inflaton (thus, breaking parity) is coupled to a vector field with a non-vanishing coherent vev (thus,  breaking statistical isotropy). We provide general expressions for the correlators as a function of the time dependence of the vector vev, under the assumption of a homogeneous coherent vev ${\bf A}^{(0)} = (0,0,A_z(\tau))$. We then evaluate these expressions in the simple case in which the amplitude decays as the square of the scale factor, as it would be obtained for the most standard ${\cal L} = - \frac{1}{4} F^2$ lagrangian for the vector field.  We show that, in this case, the resulting primordial power spectra have anisotropic quadrupolar terms with highly red-tilted ($k^{-4}$) scale dependence, and moreover, scalar-tensor and tensor-tensor power spectra violate parity. 

The $k^{-4}$ scale-dependence is due to the fact that we assume a conventional $F^2$ kinetic term for the vector field, and we impose the vector vev as an initial condition; in this case the vector vev rapidly decreases as the universe expands. This initial condition can emerge if the kinetic term of the inflaton field is of the type $f ( \phi ) F^2$, with $f$ function of the inflaton $\phi$. If the functional form of $f$ is such that $f ( \phi (t) ) \propto a^{-4}$ (where $a$ is the scale factor), then the model sustains a constant vector vev \cite{Watanabe:2009ct}. This vev arises from the unavoidable accumulation of the IR modes of the vector field, which, for $f \propto a^{-4}$, have a constant and scale invariant spectrum outside the horizon, and that therefore unavoidably add up to each other to form a coherent vev \cite{Bartolo:2012sd}.  Within this mechanism, the specific example that we study here corresponds to a transition from  $f \propto a^{-4}$ (giving constant vector vev) to $f= {\rm constant}$ (giving a decreasing vev) when the large-scale CMB modes were produced.  A constant vector vev can also emerge in models with a suitable potential $V ( \vec{A}^2 )$. One could imagine a phase transition so that the minimum of this potential changes from $\braket{\vec{A}} \neq 0$ to  $\braket{\vec{A}} = 0$ when the large-scale CMB modes were produced. Potentials that generate a vector vevs during inflation were first considered in ref.~\cite{Ford:1989me}, but they introduce ghosts \cite{Himmetoglu:2008zp}. Such a problem is not present for the $f (\phi) F^2$ mechanism. 

Computing all types of temperature and polarization power spectra, we find interesting signals that are not realized in the $f(\phi)F^2$ model or in the statistically isotropic pseudo-scalar inflation; namely, we obtain $\ell_1 = \ell_2 \pm 1$ correlations in TT, TE, EE and BB, and $\ell_1 = \ell_2 \pm 2$ in TB and EB, due to coexistence of parity violation and broken rotational invariance. Correspondence between primordial symmetry breakings and resulting CMB correlations as explained above is summarized in table~\ref{tab:correspond}.
 
This paper is organized as follows. In the next section, we summarize our model. In section~\ref{sec:primordial}, we estimate primordial scalar-scalar, scalar-tensor and tensor-tensor correlations, and in section~\ref{sec:CMB} we analyze their imprints in the CMB  (for further details, see also appendix~\ref{appen:CMB_pol}). The final section~\ref{sec:conclusion} contains our summary and discussion. 

\begin{table}[t]
\begin{center}
  \begin{tabular}{|c|c|c||c|c|c|c|c|c|c|} \hline
 Parity & Rotation & Examples & $\ell_1 = \ell_2$ & $\ell_1 = \ell_2 \pm 1$ & $\ell_1 = \ell_2 \pm 2$ \\ \hline
 $\bigcirc$ & $\bigcirc$ & Standard inflation & XX, TE & - & - \\ 
 $\times$ & $\bigcirc$ & $f(\phi)R\tilde{R}$, $f(\phi)F\tilde{F}$ & all & - & - \\
 $\bigcirc$ & $\times$ & $f(\phi)F^2$ with $\braket{\vec{A}}$ & XX, TE & TB, EB & XX, TE \\ 
 $\times$ & $\times$ & $f(\phi)F {\tilde F}$ with $\braket{\vec{A}}$  & all & all & all \\ \hline
  \end{tabular}
\end{center}
\caption{Correspondence between inflationary symmetry (parity/rotation) breakings and resulting CMB correlations. ``all'' means all possible 6 correlations: TT, TE, EE, BB, TB and EB, and XX denotes 3 auto correlations of them. We also present examples of inflationary actions creating each symmetry breaking. One can see from this table that, TT, TE, EE and BB (TB and EB) do not vanish in $\ell_1 = \ell_2 \pm 1$ ($\ell_1 = \ell_2 \pm 2$) only in the case of broken parity and rotational invariance studied in this work. 
}\label{tab:correspond}
\end{table}

\section{Pseudoscalar inflation with anisotropic gauge field} \label{sec:settings}

Let us start from an action involving an interaction between a pseudoscalar field $\phi$ and a U(1) gauge field $A_\mu$ \cite{Barnaby:2010vf}:
\begin{eqnarray}
S =  \int d^4 x \sqrt{-g} 
\left[ \frac{M_p^2}{2} R - \frac{1}{2}\partial_\mu \phi \partial^\mu \phi
- V(\phi) - \frac{1}{4} F^{\mu \nu} F_{\mu \nu} 
  -  \frac{\alpha}{4f} \phi \tilde{F}^{\mu \nu} F_{\mu \nu} \right] ~,
\end{eqnarray}
where $M_p = 1/\sqrt{8\pi G}$ is the reduced Planck mass, $R$ is the Ricci scalar, $F_{\mu \nu} \equiv \partial_\mu A_\nu - \partial_\nu A_\mu$ is the field strength, and its dual is given by $\tilde{F}^{\mu \nu} \equiv \frac{1}{2}  \frac{\eta^{\mu \nu \alpha \beta}}{\sqrt{-g}} F_{\alpha \beta}$ with $\eta^{0123} \equiv 1$. The coupling strength between the inflaton $\phi$ and the vector field is given by  $\alpha / f$, where $f$ is the inflaton decay constant and $\alpha$ a dimensionless parameter that is naturally expected to be of order one. We assume that the potential  $V(\phi)$ sustains  slow-roll inflation. 

The previous studies \cite{Barnaby:2010vf, Barnaby:2011vw, Sorbo:2011rz} examined cosmological phenomena induced by the gauge field, e.g., parity-violating gravitational waves or large equilateral non-Gaussianity of curvature perturbations, assuming isotropic evolution of the background Universe. On the other hand, as a new point view of this work, we want to investigate impacts in perturbed quantities if the gauge field has a nonzero vacuum expectation value (vev). One may be concerned that, in such case, a nontrivial direction dependence of the gauge field causes an anisotropic background evolution. However, in analogy with refs.~\cite{Watanabe:2010fh, Bartolo:2012sd, Maleknejad:2013npa}, such anisotropy is rapidly smoothed in the inflationary expansion, if the energy density of the gauge field is much smaller than the pseudoscalar energy density:
\begin{eqnarray}
\rho_\phi \simeq V  \simeq 3 M_p^2 H^2 \gg \frac{1}{2} \left( E^2 + B^2 \right)~. \label{eq:potential}
\end{eqnarray}
Here, $H$ is the Hubble parameter and we have used the electromagnetic decomposition as ${\bf E} \equiv - {\bf A}' / a^2$ and ${\bf B} \equiv \nabla \times {\bf A} / a^2$ on the Coulomb gauge, with $~'~ \equiv \partial_\tau$ denoting the derivative of the conformal time $\tau$. In the following discussion, we follow the condition \eqref{eq:potential}, and hence work on the isotropic background metric: $ds^2 = a^2(\tau)( -d\tau^2 + dx^2)$. 
Therefore, in the limit in which the vev of the gauge field goes to zero we recover the statistically isotropic results of \cite{Barnaby:2010vf, Barnaby:2011vw, Sorbo:2011rz}. For small but non-vanishing vev we expect that small anisotropic modulations appear in primordial correlations via the $\phi F {\tilde F}$ interaction, imprinting  new signatures in the CMB correlation functions. 

In our scenario, the gauge field may be divided into the background and perturbed parts as $A_\mu(\tau, {\bf x}) \equiv  A_\mu^{(0)}(\tau, {\bf x}) + \delta A_\mu(\tau, {\bf x})$. For convenience, let us choose the Coulomb gauge: $A_0 = 0$ and $\nabla \cdot {\bf A} = 0$, without loss of generality. Then, the equation of motion for the gauge field reads \cite{Barnaby:2011vw}
\begin{eqnarray}
{\bf A}'' - \nabla^2 {\bf A} + \frac{2 \xi}{\tau} \nabla \times {\bf A}
= 0 \label{eq:EOM_vec}
\end{eqnarray}
where $\xi \equiv \frac{\alpha \dot{\phi}}{2fH}$ with $~\dot{}~ \equiv \partial_t$ denoting the derivative of the physical time.

In principle, one may consider various configurations for the  background gauge field. In this work we consider the extremely simple case in which  ${\bf A}^{(0)} = (0, 0, A_z(\tau))$. Substituting this into eq.~\eqref{eq:EOM_vec} yields the usual $a^{-2}$-decaying electric field and no magnetic field:
\begin{eqnarray}
{\bf E}^{(0)}(\tau) 
= {\bf E}^{\rm in} \left(\frac{\tau}{\tau_{\rm in}}\right)^2
\simeq {\bf E}^{\rm in} \left(\frac{a_{\rm in}}{a}\right)^2
 ~, \ \ 
{\bf B}^{(0)}(\tau) = 0 ~, \label{eq:BG_vec}
\end{eqnarray}
where ${\bf E}^{\rm in}$ is the electric field at a given initial time and we have used $a \simeq -(H \tau)^{-1}$.

The fluctuation part is expanded with a divergenceless polarization vector $\epsilon_i^{(\pm 1)}$ (defined in appendix~\ref{appen:pol}) as 
\begin{eqnarray}
 \delta A_i(\tau, {\bf x}) \equiv 
\int \frac{d^3{\bf k}}{(2\pi)^{3/2}}
\sum_{\lambda = \pm 1} 
\delta A_{\bf k}^{(\lambda)}(\tau) 
\epsilon_i^{(\lambda)}({\bf k})
 e^{i {\bf k} \cdot {\bf x}} ~.
\end{eqnarray}
The gauge field is quantized by the creation ($a_\lambda^\dagger$) and annihilation ($a_\lambda$) operators:  
\begin{eqnarray}
\delta \hat{A}_{\bf k}^{(\lambda)}(\tau) 
&=& a_{\lambda}({\bf k}) \delta A_\lambda (\tau, k) 
+ a_{\lambda}^\dagger(- {\bf k}) \delta A_\lambda^* (\tau, k)  ~, 
\end{eqnarray}
with the algebra $[a_{\lambda}({\bf k}), a_{\lambda'}^\dagger({\bf k}')] = \delta_{\lambda \lambda'} \delta^{(3)}({\bf k} - {\bf k}')$. This decomposition reduces eq.~\eqref{eq:EOM_vec} to an equation in terms of the mode function $\delta A_\lambda$~\cite{Anber:2006xt}: 
\begin{eqnarray}
\left[ \partial_\tau^2 + k^2 + \frac{2 \lambda k \xi}{\tau} \right] \delta A_\lambda(\tau,k)  = 0 ~.
\end{eqnarray}
Choosing $\dot{\phi} > 0$ (and hence $\xi > 0$) during inflation without loss of generality, we obtain the solutions including a growing $\lambda = +$ mode coming from tachyonic instability and a negligible $\lambda = -$ one: \cite{Anber:2006xt,Barnaby:2010vf, Barnaby:2011vw, Sorbo:2011rz}  
\begin{eqnarray}
\delta A_+(\tau, k) \simeq \frac{1}{\sqrt{2k}} \left( - \frac{k\tau}{2 \xi} \right)^{1/4} e^{\pi\xi - 2\sqrt{- 2\xi k \tau}} ~, \ \ 
|\delta A_-(\tau, k)| \ll |\delta A_+(\tau, k)| ~,
\label{dA-sol}
\end{eqnarray}
in the interval $(8\xi)^{-1} \lesssim -k\tau \lesssim 2\xi$. For $\xi \gtrsim 1$, the contribution of the growing mode becomes significant because of the exponential enhancement $e^{\pi \xi}$. Due to this strongly polarized nature, the correlations of the electric and magnetic fields are given by the + mode alone:
\begin{eqnarray}
\Braket{\delta X_i(\tau, {\bf k}) \delta Y_j(\tau', {\bf k}')} 
\approx \delta X(\tau, k) \delta Y(\tau', k)
 \epsilon_i^{(+)}(\hat{\bf k}) \epsilon_j^{(+)}(\hat{\bf k}') 
\delta^{(3)}({\bf k} + {\bf k}')~, \label{eq:pow_vec}
\end{eqnarray}
where $X, Y = E, B$ and 
\begin{eqnarray}
\delta E(\tau, k) =   
 - H^2 \tau^2 \sqrt{k}
 \left( - \frac{\xi}{2k\tau} \right)^{1/4} e^{\pi\xi - 2\sqrt{- 2\xi k \tau}} 
\approx - \sqrt{\frac{2\xi}{|k\tau|}} \delta B(\tau, k)
~. \label{eq:deltaEB}
\end{eqnarray}

\section{Broken parity and rotational invariance in primordial power spectra} \label{sec:primordial}

In our scenario, in addition to the usual isotropic vacuum contributions, power spectra of curvature perturbations and gravitational waves also acquire corrections originating from the interactions with the gauge field: 
\begin{eqnarray}
{\cal L}_{\rm int} =  - \frac{1}{4} F^{\mu \nu} F_{\mu \nu} 
  -  \frac{\alpha}{4f} \phi \tilde{F}^{\mu \nu} F_{\mu \nu} ~. \label{eq:lag3}
\end{eqnarray}
The direction dependence of the background gauge field and parity-violating features in the gauge field fluctuations are directly reflected in the curvature and gravitational wave power spectra. 
In this section, we compute them by means of the in-in formalism \cite{Maldacena:2002vr,Weinberg:2005vy}. 

In the computation of the cosmological perturbations, we follow the same conventions used in previous literature \cite{Barnaby:2011vw, Barnaby:2012xt}. We work with the Arnowitt-Deser-Misner (ADM) form of the metric and adopt the spatially flat gauge. The spatial perturbations of the metric are then generated by the tensor perturbation modes alone, reading $\delta g_{ij} = a^2 h_{ij}$. The pseudoscalar field is decomposed into the background and perturbative parts: $\phi = \phi^{(0)} + \delta \phi$. The $\delta g_{00}$ and $\delta g_{0i}$ modes can be expressed in terms of $\delta \phi$, $\delta A_i$ and $h_{ij}$ by use of the so-called energy and momentum constraints. 

Expanding the interactions \eqref{eq:lag3} with curvature perturbations $\zeta \simeq - \frac{H}{\dot{\phi}} \delta \phi$, $\delta A_i$ and $h_{ij}$ yields the tree-level interaction Hamiltonian:
\begin{eqnarray}
H_1(\tau)
= \frac{E_i^{\rm in}}{H^4 \tau_{\rm in}^2 \tau^2}  
\int d^3 {\bf p}  \left[  
2 \xi \delta B_i(\tau, {\bf p}) 
\hat{\zeta}_{- {\bf p}}(\tau)  
+ \delta E_j(\tau, {\bf p}) 
\hat{h}_{ij, - {\bf p}}(\tau)
\right]~. \label{eq:Hint}
\end{eqnarray}
Here, we have dropped the interactions between the $F^2$ term and the scalar curvature perturbation $\zeta$, since these are suppressed by the slow-roll parameters and turn out to be much smaller than the contributions from the $\phi F\tilde{F}$ term with $\xi = {\cal O}(1)$ \cite{Barnaby:2011vw, Barnaby:2012xt}. On the other hand, the only coupling between the vector field and the tensor part if given by  the $F^2$ term, 
giving ride to the second term in $H_1$. 

Curvature perturbations and gravitational waves are expanded as 
\begin{eqnarray}
\zeta(\tau, {\bf x}) &=& \int \frac{d^3{\bf k}}{(2\pi)^{3/2}}
\zeta_{\bf k}(\tau) 
 e^{i {\bf k} \cdot {\bf x}} ~, \\
h_{ij}(\tau, {\bf x}) &=& 
\int \frac{d^3{\bf k}}{(2\pi)^{3/2}}
\sum_{\lambda = \pm 2} 
h_{\bf k}^{(\lambda)}(\tau) 
e_{ij}^{(\lambda)}({\bf k})
 e^{i {\bf k} \cdot {\bf x}} ~, 
\end{eqnarray}
where $e_{ij}^{(\pm 2)}$ is a transverse and traceless polarization tensor defined in appendix~\ref{appen:pol}. 
Their power spectra are composed of the vacuum contribution (0 mode) and the tree-level correction coming from the interaction (1 mode) (we note that the vacuum mode is uncorrelated with the mode arising from the interaction \eqref{eq:Hint}):
\begin{eqnarray} 
  \Braket{\zeta_{{\bf k}_1} \zeta_{{\bf k}_2} } &=& \Braket{\zeta_{{\bf k}_1} \zeta_{{\bf k}_2}}_0 + \Braket{\zeta_{{\bf k}_1} \zeta_{{\bf k}_2}}_1 ~, \\
  \Braket{h_{{\bf k}_1}^{(\lambda_1)} h_{{\bf k}_2}^{(\lambda_2)}} &=& \Braket{h_{{\bf k}_1}^{(\lambda_1)} h_{{\bf k}_2}^{(\lambda_2)}}_0 + \Braket{h_{{\bf k}_1}^{(\lambda_1)} h_{{\bf k}_2}^{(\lambda_2)}}_1  ~, \\ 
  \Braket{\zeta_{{\bf k}_1} h_{{\bf k}_2}^{(\lambda_2)}} &=&  \Braket{\zeta_{{\bf k}_1}  h_{{\bf k}_2}^{(\lambda_2)}}_1 ~.
\end{eqnarray}
Note that $\Braket{\zeta_{{\bf k}_1} h_{{\bf k}_2}^{(\lambda_2)}}_0 = 0$ because mode couplings are only allowed in anisotropic correlations.

The vacuum modes are  quantized in terms of  the spin-0 and 2 creation and annihilation operators 
\begin{eqnarray}
\hat{\zeta}_{\bf k}(\tau) 
&=& a_{0}({\bf k}) \zeta(\tau, k) 
+ a_{0}^\dagger(- {\bf k}) \zeta^* (\tau, k) ~, \\
\hat{h}_{\bf k}^{(\lambda)}(\tau) 
&=& a_{\lambda}({\bf k}) h (\tau, k) 
+ a_{\lambda}^\dagger(- {\bf k}) h^* (\tau, k) ~,
\end{eqnarray}
with $[a_{\lambda}({\bf k}), a_{\lambda'}^\dagger({\bf k}')] = \delta_{\lambda, \lambda'} \delta^{(3)}({\bf k} - {\bf k}')$ for $\lambda, \lambda' = 0, \pm 2$. Solving the equations of motion coming from the quadratic actions with respect to $\delta \phi$ and $h_{ij}$ yields the vacuum mode functions at leading order in slow roll, namely
\begin{eqnarray}
\zeta(\tau, k) =  \frac{h(\tau, k)}{2\sqrt{\epsilon}} \simeq
\frac{i H (1 + ik\tau)}{2\sqrt{\epsilon} M_p k^{3/2}} e^{-ik\tau}  ~,
\end{eqnarray}
where $\epsilon = \frac{1}{2} \left( \frac{\dot{\phi}}{H M_p} \right)^2$ is a slow-roll parameter for $\phi$. These mode functions determine the 0-mode power spectra as 
\begin{eqnarray}
\begin{split}
  \Braket{\zeta_{{\bf k}_1} \zeta_{{\bf k}_2} }_0 &= 
\frac{2\pi^2}{k_1^3} {\cal P} \delta^{(3)}({\bf k}_1 + {\bf k}_2) ~, \\
  \Braket{h_{{\bf k}_1}^{(\lambda_1)} h_{{\bf k}_2}^{(\lambda_2)}}_0
&= \frac{8\pi^2}{k_1^3} \epsilon {\cal P} 
\delta^{(3)}({\bf k}_1 + {\bf k}_2) \delta_{\lambda_1, \lambda_2} 
 ~.
\end{split}
\label{eq:pow_ini_0}
\end{eqnarray}
where ${\cal P} = \frac{H^2}{8\pi^2 \epsilon M_p^2}$. The tree-level 1-mode correlations are computed in the following subsections.

\subsection{Scalar-scalar correlation}

According to the in-in formalism, the tree-level (1-mode) correlation of curvature perturbations at given $\tau$ is expressed as 
\begin{eqnarray}
\Braket{ \zeta_{{\bf k}_1} \zeta_{{\bf k}_2} (\tau) }_1 
&=& - \int_{\tau_{\rm in}}^{\tau} d\tau_1
\int_{\tau_{\rm in}}^{\tau_1} d\tau_2
\Braket{ \left[\left[ \hat{\zeta}_{{\bf k}_1} \hat{\zeta}_{{\bf k}_2}(\tau), H_1(\tau_1) \right], H_1(\tau_2) \right]
} ~.
\end{eqnarray}
 Since the vacuum scalar and tensor mode functions are uncorrelated, the tensor part in the interaction Hamiltonian \eqref{eq:Hint} is not relevant to this computation. Moreover, the produced gauge modes are classical (specifically, we see that the gauge field  has vanishing commutators when the mode function is given by (\ref{dA-sol}), therefore the effect that we are computing is proportional to the classical produced gauge modes), and hence only the curvature perturbations have a nontrivial commutator, giving 
\begin{eqnarray} 
\Braket{ \zeta_{{\bf k}_1} \zeta_{{\bf k}_2} (\tau) }_1  
&=& - \frac{ 4 \xi^2 E_{\rm in}^2  }{H^8 \tau_{\rm in}^4} \int_{\tau_{\rm in}}^{\tau} \frac{d\tau_1}{\tau_1^2} 
\int_{\tau_{\rm in}}^{\tau_1} \frac{d\tau_2}{\tau_2^2}
\int d^3 {\bf p}_1 \int d^3 {\bf p}_2   \nonumber \\ 
&&\times 
 \hat{E}_i^{\rm in} \hat{E}_j^{\rm in}
\Braket{ \delta B_i(\tau_1, {\bf p}_1) \delta B_j(\tau_2, {\bf p}_2)  } \nonumber \\ 
&&\times 
\left( 
\Braket{ \left[ \hat{\zeta}_{{\bf k}_1}(\tau) , \hat{\zeta}_{- {{\bf p}_1}}(\tau_1)  \right] } 
\Braket{ \left[ \hat{\zeta}_{{\bf k}_2}(\tau) , \hat{\zeta}_{- {{\bf p}_2}}(\tau_2) \right] }  + ({\bf k}_1 \leftrightarrow {\bf k}_2)
\right)  
~,
\end{eqnarray}
where $E_{\rm in} \equiv |{\bf E}^{\rm in}|$. Computing the correlations with eq.~\eqref{eq:pow_vec} and  
\begin{eqnarray}
\Braket{ \left[ \hat{\zeta}_{\bf k}(\tau), \hat{\zeta}_{{\bf k}'}(\tau') \right] }
 = \left( \zeta(\tau, k) \zeta^*(\tau', k) - c.c. \right)
\delta^{(3)}({\bf k} + {\bf k}' ) ~,
\end{eqnarray}
and using the identity
\begin{eqnarray}
 \sum_{s= \pm 1} 
\epsilon_i^{(s)}(\hat{\bf k}) \epsilon_j^{(-s)}(\hat{\bf k})
 \hat{E}_i^{\rm in} \hat{E}_j^{\rm in} 
= 1 - \left( \hat{\bf k} \cdot \hat{\bf E}^{\rm in} \right)^2  
~, \label{eq:prod_eps_E}
\end{eqnarray}
 we can reach a reduced form:
\begin{eqnarray} 
\Braket{ \zeta_{{\bf k}_1} \zeta_{{\bf k}_2} (\tau) }_1 
=  \left[ 1 - \left( \hat{\bf k}_1 \cdot \hat{\bf E}^{\rm in} \right)^2 \right]
 f_{\zeta \zeta}(k_1,\tau) 
\delta^{(3)}({\bf k}_1 + {\bf k}_2)
~, \label{eq:zeta_zeta_1}
\end{eqnarray}
with 
\begin{eqnarray}
f_{\zeta \zeta}(k,\tau)  
&\equiv& 
- \frac{2 \xi^2 E_{\rm in}^2}{H^8 \tau_{\rm in}^4} 
\left[ \int_{\tau_{\rm in}}^{\tau} 
\frac{d\tau_1 }{\tau_1^2}  
\delta B(\tau_1, k) 
\left( \zeta(\tau, k) \zeta^*(\tau_1, k) - c.c. \right) \right]^2 ~.
 \label{eq:f_zeta_zeta}
\end{eqnarray}
In the derivation of eq.~\eqref{eq:f_zeta_zeta}, on the basis of the symmetric property of the integrand under $\tau_1 \leftrightarrow \tau_2$, we have changed the integral interval as $\int_{\tau_{\rm in}}^{\tau} d\tau_1 \int_{\tau_{\rm in}}^{\tau_1} d\tau_2 = \frac{1}{2}\int_{\tau_{\rm in}}^{\tau} d\tau_1 \int_{\tau_{\rm in}}^{\tau} d\tau_2$. 
This expression shows that the tree-level correction creates a monopolar contribution and a quadrupolar modulation in the curvature power spectrum.

We are now interested in the correlation on superhorizon scales $-k\tau \ll 1$. In such limit, the time integrals are also determined by the superhorizon contributions, hence we can evaluate them as 
\begin{eqnarray}
f_{\zeta \zeta}(k,\tau)  
&\simeq& 
\frac{4 \xi^2 E_{\rm in}^2}{12^2 \epsilon^2 M_p^4 \tau_{\rm in}^4 k^7}
\left( \frac{1}{2 \xi} \right)^{1/2}
 e^{2 \pi\xi} 
\left[\int^{-k \tau_{\rm in}}_{-k \tau} 
dx x^{13/4} e^{- 2\sqrt{2\xi x}} \right]^2 ~,
\end{eqnarray}
where we have adopted an approximation: $\zeta(\tau, k) \zeta^*(\tau_1, k) - c.c.
\simeq - \frac{i H^2}{6\epsilon M_p^2} \left( \tau^3 - \tau_1^3 \right)$. If $\xi = {\cal O}(1)$, the integrand has a peak at around $x \sim \xi$ and rapidly decays for both $x \ll \xi$ and $x \gg \xi$. Therefore, assuming $-k \tau \ll \xi \ll -k \tau_{\rm in}$, we can safely regard the interval of the time integral as $\int^{-k \tau_{\rm in}}_{-k\tau} dx \to \int^{\infty}_{0} dx$. Such integral is analytically solved as 
\begin{eqnarray}
\int^{\infty}_0 dx x^{n} e^{- z \sqrt{x}} 
&=& \frac{2 \Gamma\left(2n+2\right)}{z^{2n+2}} \ \ (n > -1)  ~,
\end{eqnarray}
and thus we can obtain the superhorizon expression:
\begin{eqnarray}
f_{\zeta \zeta}(k, \tau) 
\simeq 
\frac{\Gamma^2(\frac{17}{2})}{9 \times  2^{26}} 
\frac{e^{2 \pi\xi}}{\xi^7} 
\frac{E_{\rm in}^2}{\epsilon^2 M_p^4 \tau_{\rm in}^4 k^7} 
~.
\end{eqnarray}
It is verified from this result that the monopolar contribution and the quadrupolar modulation correlator scale as  $k^{-7}$ due to the $a^{-2}$ decaying feature of the background gauge field.

\subsection{Tensor-tensor correlation}

The auto correlation of gravitational waves is given by the in-in formalism:
\begin{eqnarray}
\Braket{h_{{\bf k}_1}^{(\lambda_1)} h_{{\bf k}_2}^{(\lambda_2)}(\tau)}_1 
&=& - \int_{\tau_{\rm in}}^{\tau} d\tau_1
\int_{\tau_{\rm in}}^{\tau_1} d\tau_2
\Braket{ \left[\left[ \hat{h}_{{\bf k}_1}^{(\lambda_1)} \hat{h}_{{\bf k}_2}^{(\lambda_2)} (\tau), H_1(\tau_1) \right], H_1(\tau_2) \right]
} ~.
\end{eqnarray}
We can compute it in the same manner as the scalar case, on the other hand only the gravitational wave part in the interaction Hamiltonian contributes and hence $\Braket{\delta E_i \delta E_{j}}$ is involved. We finally obtain a reduced form with the angle dependence identical to the scalar case:
\begin{eqnarray}
\Braket{h_{{\bf k}_1}^{(\lambda_1)} h_{{\bf k}_2}^{(\lambda_2)}(\tau)}_1 
= \left[1 - \left( \hat{\bf k}_1 \cdot \hat{\bf E}^{\rm in} \right)^2 \right] 
 f_{hh}(k_1,\tau) 
 \delta^{(3)}({\bf k}_1 + {\bf k}_2) \delta_{\lambda_1, 2} \delta_{\lambda_2, 2}
~. \label{eq:h_h_1}
\end{eqnarray}
The $\lambda = +2$ polarized feature in this form due to eq.~\eqref{eq:pow_vec} creates interesting CMB correlations associated with the parity violation, as discussed in the next section. The amplitude of the 1-mode power spectrum is relevant to the electric field alone since there is no magnetic-field contribution in the tensor part of the tree-level interaction Hamiltonian \eqref{eq:Hint}. 

The approximations for $-k\tau \ll \xi \ll -k \tau_{\rm in}$, like the above scalar case, yield a superhorizon expression of the radial function:
\begin{eqnarray}
f_{hh}(k, \tau) 
&\equiv& - \frac{E_{\rm in}^2}{H^8 \tau_{\rm in}^4} 
\left[ \int_{\tau_{\rm in}}^\tau \frac{d\tau_1}{\tau_1^2} 
 \delta E(\tau_1, k) 
\left( h(\tau,k) h^*(\tau_1, k) - c.c. \right) \right]^2
\nonumber \\ 
&\simeq& \frac{\Gamma^2(\frac{15}{2})}{9 \times 2^{19}}  
 \frac{e^{2 \pi\xi} }{\xi^7} 
\frac{E_{\rm in}^2 }{M_p^4 \tau_{\rm in}^4 k^7}
~,
\end{eqnarray}
where we have used
\begin{eqnarray}
\Braket{ \left[ \hat{h}^{(\lambda)}_{\bf k}(\tau), \hat{h}^{(\lambda')}_{{\bf k}'}(\tau') \right] }
 &=& \left( h(\tau, k) h^*(\tau', k) - c.c. \right)
 \delta^{(3)}({\bf k} + {\bf k}') \delta_{\lambda, \lambda'} \nonumber \\ 
&\simeq & 
- \frac{2 i H^2}{3 M_p^2} 
\left( \tau^3 - \tau'^3 \right) \delta^{(3)}({\bf k} + {\bf k}') \delta_{\lambda, \lambda'}  ~.
\end{eqnarray}
Notice again the $k^{-7}$ dependence of the 1-mode correlator. 

\subsection{Scalar-tensor correlation}

The cross correlation between curvature perturbations and gravitational waves can be formulated as 
\begin{eqnarray}
\Braket{\zeta_{{\bf k}_1} h_{{\bf k}_2}^{(\lambda_2)}(\tau)}_1 
&=& - \int_{\tau_{\rm in}}^{\tau} d\tau_1
\int_{\tau_{\rm in}}^{\tau_1} d\tau_2
\Braket{ \left[ \left[ \hat{\zeta}_{{\bf k}_1}
\hat{h}_{{\bf k}_2}^{(\lambda_2)}(\tau), H_{1}(\tau_1) \right] , H_{1}(\tau_2) \right] } ~.
\end{eqnarray}
A non-vanishing result is now obtained from the coupling between the two terms of the integration Hamiltonian, eq.~(\ref{eq:Hint}), namely when $H_1^{B \zeta} \left( \tau_1 \right)$ is considered together with 
$H_1^{Eh} \left( \tau_2 \right)$ (or viceversa).  The result is thus proportional to the  $\Braket{\delta B_i \delta E_{j}}$ correlator. Proceeding similarly to the  auto correlation cases, we obtain 
\begin{eqnarray}
\Braket{\zeta_{{\bf k}_1} h_{{\bf k}_2}^{(\lambda_2)}(\tau)}_1 
=  
\hat{E}_i^{\rm in} \hat{E}_j^{\rm in} e_{ij}^{(\lambda_2)}(\hat{\bf k}_1) 
 f_{\zeta h}(k_1,\tau) 
 \delta^{(3)}({\bf k}_1 + {\bf k}_2) \delta_{\lambda_2, 2}
~. \label{eq:zeta_h_1}
\end{eqnarray}
The contraction with $e_{ij}^{(\lambda)}(\hat{\bf k})$ is generally expanded with the spin spherical harmonics ${}_{\lambda}Y_{2M}(\hat{\bf k})$, as shown in eq.~\eqref{eq:identity_e2EE}. If we choose ${\bf E}^{\rm in}$ to be parallel to the $z$ axis, namely $\hat{\bf E}^{\rm in} = (0,0,1)$, the contraction is simplified as
\begin{eqnarray}
\hat{E}_i^{\rm in} \hat{E}_j^{\rm in} e_{ij}^{(\lambda)}(\hat{\bf k}) 
= \frac{1}{\sqrt{2}} 
\left[ 1 - \left(\hat{\bf k} \cdot \hat{\bf E}^{\rm in}\right)^2 \right]
~. \label{eq:angle_z_zeta_h}
\end{eqnarray}
These explicitly show that the quadrupolar modulation is realized also in the scalar-tensor correlation.

Considering the superhorizon limit in the same manner as the auto correlation cases, they are evaluated as
\begin{eqnarray}
 f_{\zeta h}(k, \tau) 
&\equiv& - \frac{2 \xi E_{\rm in}^2}{H^8 \tau_{\rm in}^4}
\left[  \int_{\tau_{\rm in}}^{\tau} \frac{d\tau_1}{\tau_1^2} 
\delta B(\tau_1, k) \left( \zeta(\tau, k) \zeta^*(\tau_1, k) - c.c. \right) 
\right] \nonumber \\ 
&&\times \left[  \int_{\tau_{\rm in}}^{\tau} \frac{d\tau_2}{\tau_2^2} 
\delta E(\tau_2, k) \left( h(\tau, k) h^*(\tau_2, k) - c.c. \right) 
 \right]
\nonumber \\ 
&\simeq&
- \frac{\Gamma(\frac{15}{2}) \Gamma(\frac{17}{2})}{9 \times 2^{22}} 
\frac{e^{2 \pi\xi}}{\xi^7} 
\frac{E_{\rm in}^2 }{\epsilon M_p^4 \tau_{\rm in}^4 k^7} 
  ~.
\end{eqnarray}
Note that the correlation is negative, owing to the difference of the sign between $\delta B$ and $\delta E$, as seen in eq.~\eqref{eq:deltaEB}.

\section{Broken parity and rotational invariance in CMB power spectra} \label{sec:CMB}

In this section we analyze the CMB signatures induced by the anisotropic scalar-scalar~\eqref{eq:zeta_zeta_1}, tensor-tensor~\eqref{eq:h_h_1} and scalar-tensor~\eqref{eq:zeta_h_1} correlations. 
We start with the discussion on the temperature power spectra, and then explain how non-vanishing signals in the distinct $\ell$-space configurations associated with parity-violating statistical anisotropy, i.e., $\ell_1 = \ell_2 \pm 1$ arise (see table~\ref{tab:correspond}). Comprehensive analysis including the other special correlations: TE, EE and BB in $\ell_1 = \ell_2 \pm 1$, and TB and EB in $\ell_1 = \ell_2 \pm 2$ is presented in appendix~\ref{appen:CMB_pol}.

In the computation of the CMB power spectra, for convenience, we utilize the spherical harmonic representations of the initial 1-mode power spectra \eqref{eq:zeta_zeta_1}, \eqref{eq:h_h_1} and \eqref{eq:zeta_h_1}:
\begin{eqnarray}
\Braket{\zeta_{{\bf k}_1} \zeta_{{\bf k}_2}}_1 
&=&
 \frac{2\pi^2 {\cal P}}{k_1^3}
\left( \frac{k_0}{k_1}\right)^4 
\sum_{L = 0,2} \sum_{M} g_{LM}^{ss} 
Y_{LM} (\hat{\bf k}_1) 
\delta^{(3)}({\bf k}_1 + {\bf k}_2)
~, \label{eq:zeta_zeta_1_sYlm} \\ 
\Braket{h_{{\bf k}_1}^{(\lambda_1)} h_{{\bf k}_2}^{(\lambda_2)}}_1 
&=& 
\frac{2\pi^2 {\cal P}}{k_1^3}
\left( \frac{k_0}{k_1}\right)^4 \sum_{L = 0,2} \sum_{M}
g_{LM}^{tt} Y_{LM} (\hat{\bf k}_1) 
\delta^{(3)}({\bf k}_1 + {\bf k}_2)
\delta_{\lambda_1, 2}
 \delta_{\lambda_2, 2}
~ , \label{eq:h_h_1_sYlm} \\ 
\Braket{ \zeta_{{\bf k}_1} h_{{\bf k}_2}^{(\lambda_2)} }_1 
&=&  \frac{2\pi^2 {\cal P}}{k_1^3}
\left( \frac{k_0}{k_1}\right)^4 \sum_{M} 
g_{2M}^{st}\,\,
{}_{\lambda_2}Y_{2M} (\hat{\bf k}_1) 
\delta^{(3)}({\bf k}_1 + {\bf k}_2)
\delta_{\lambda_2, 2}
~, \label{eq:zeta_h_1_sYlm}
\end{eqnarray}
where $1/k_0 = 14~{\rm Gpc}$ is a pivot scale comparable to the present horizon scale, and we have followed the rules for harmonic expansion described in appendix~\ref{appen:pol}.
\footnote{
Notice that the pseudoscalar inflation model we are considering provides an example of a strong scale dependent quadrupolar modulation, with, in the notations of ref.~\cite{Ma:2011ii}, $g_{LM}(k)=g_{LM}(0.002~{\rm Mpc^{-1}} /k)^q$ with $q=4$, while here we adopt another pivot scale and consider also tensor perturbation modes.} 
The scalar monopolar coefficient $g_{00}^{ss}$ depends on the energy density of the gauge field when the CMB scales ($k_0$) leave the horizon during inflation, $\rho_E^{\rm CMB} \equiv E_{\rm CMB}^2 / 2 = \frac{E_{\rm in}^2}{2 (k_0 \tau_{\rm in})^4} $, while the quadrupolar coefficient $g_{2M}^{ss}$ involves both $\rho_E^{\rm CMB}$ and the direction dependence $\hat{\bf E}^{\rm in}$:
\begin{eqnarray}
\begin{split}
g_{00}^{ss} &\simeq  
 \frac{4\sqrt{\pi}}{3} 
\frac{\Gamma^2\left(\frac{17}{2}\right)}{12^2 2^{22}} \frac{e^{2 \pi\xi}}{\xi^7}
\frac{\rho_E^{\rm CMB}}{\epsilon^2 \pi^2 {\cal P} M_p^4}\, , \\
g_{2M}^{ss} 
& \simeq - \frac{2\sqrt{\pi}}{5} g_{00}^{ss}
Y_{2M}^* (\hat{\bf E}^{\rm in}) ~. 
\end{split}
\label{eq:gLM_def}
\end{eqnarray}
The coefficients of the tensor-tensor or scalar-tensor power spectrum are suppressed by $\epsilon^2$ or $\epsilon$ in comparison with $g_{LM}^{ss}$, namely (at leading order in the slow-roll parameters)
\begin{eqnarray}
g_{LM}^{tt} \simeq 2 \left( \frac{16}{15}\right)^2 \epsilon^2 g_{LM}^{ss} ~, \ \
g_{2M}^{st} \simeq \frac{32\sqrt{3}}{15} \epsilon g_{2M}^{ss} ~. \label{eq:consistency}
\end{eqnarray}
Note that the auto power spectra have both monopolar ($L=0$) and quadrupolar ($L=2$) components, while the cross power spectrum is given only by a quadrupolar component. In the final part of this section, we estimate the expected uncertainties on $g_{LM}^{ss}$ by means of a Fisher matrix analysis.

\begin{figure}[t]
  \begin{tabular}{c}
    \begin{minipage}{1.0\hsize}
  \begin{center}
    \includegraphics[width = 1\textwidth]{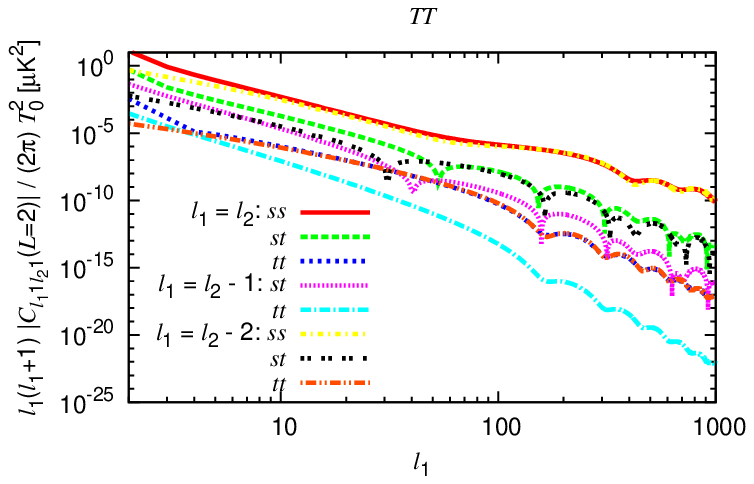}
  \end{center}
\end{minipage}
\end{tabular}
  \caption{Anisotropic (1-mode) temperature power spectra $C_{\ell_1 1 \ell_2 1}^{ss;TT}(L = 2)$, $C_{\ell_1 1 \ell_2 1}^{st;TT}(L = 2)$ and $C_{\ell_1 1 \ell_2 1}^{tt;TT}(L = 2)$,  for $g_{20}^{ss} = 1$, ${\cal P} = 2.5 \times 10^{-9}$ and $\epsilon = 0.01$. Here, we fix the direction of the initial gauge field to be $\hat{\bf E}^{\rm in} = (0,0,1)$ and hence $Y_{2M}(\hat{\bf E}^{\rm in}) = \sqrt{\frac{5}{4\pi}}\delta_{M,0}$ holds. Non-vanishing parity-odd ($\ell_1 = \ell_2 - 1$) signals in the scalar-tensor and tensor-tensor power spectra, and expected $\ell^{-4}$ scaling due to the initial power spectra proportional to $k^{-7}$ are confirmed.} \label{fig:Cl_II}
\end{figure}

\subsection{Scalar-scalar correlation}

The harmonic representation of the CMB temperature anisotropy of the scalar mode is given by 
\begin{eqnarray}
a_{T,\ell m}^{(s)} = 
4\pi (-i)^{\ell} \int \frac{d^3 {\bf k}}{(2\pi)^{3/2}}
 {\cal T}_{T,\ell}^{(s)}(k) \zeta_{\bf k} Y_{\ell m}^*(\hat{\bf k})~. \label{eq:alm_S}
\end{eqnarray}
where ${\cal T}_{T,\ell}^{(s)}(k)$ is the radiation transfer function of scalar temperature modes. The 0-mode power spectrum generated from eq.~\eqref{eq:pow_ini_0} is given by the usual rotational-invariant form: 
\begin{eqnarray}
\Braket{a_{T,\ell_1 m_1}^{(s)} a_{T,\ell_2 m_2}^{(s) *}}_0 
=  4\pi  {\cal P}
\int_0^\infty \frac{dk}{k} \left[ {\cal T}_{T,\ell_1}^{(s)}(k) \right]^2
 \delta_{\ell_1, \ell_2} \delta_{m_1, m_2} ~.
\end{eqnarray}

Using the harmonic representation of the curvature power spectrum~\eqref{eq:zeta_zeta_1_sYlm}, the CMB 1-mode power spectrum is formulated as
\begin{eqnarray}
\Braket{a_{T,\ell_1 m_1}^{(s)} a_{T,\ell_2 m_2}^{(s) *}}_1
&=& i^{\ell_2 -\ell_1}
4\pi  {\cal P}
\int_0^\infty \frac{dk}{k}
{\cal T}_{T,\ell_1}^{(s)}(k)  
{\cal T}_{T,\ell_2}^{(s)}(k) \left(\frac{k_0}{k}\right)^4
\nonumber \\ 
&&\times 
\sum_{L = 0,2} \sum_{M}  g_{LM}^{ss}
\int d^2 \hat{\bf k} Y_{\ell_1 m_1}^*(\hat{\bf k})
Y_{\ell_2 m_2}(\hat{\bf k}) 
Y_{LM} (\hat{\bf k}) \nonumber \\
&\equiv& \sum_{L=0,2} C_{\ell_1 m_1 \ell_2 m_2}^{ss;TT}(L) ~.
\end{eqnarray}
The angular integral determining $\ell$-space configurations can be solved by using the Gaunt integral formula:
\begin{eqnarray}
\label{eq:gaunt}
\int d^2 {\bf n} {}_{s_1}Y_{l_1 m_1} {}_{s_2}Y_{l_2 m_2}  {}_{s_3}Y_{l_3 m_3} 
&=& I^{-s_1 -s_2 -s_3}_{~l_1~ l_2~ l_3}  
\left(
  \begin{array}{ccc}
  l_1 & l_2 & l_3 \\
  m_1 & m_2 & m_3
  \end{array}
 \right) ~, \\
I^{s_1 s_2 s_3}_{l_1 l_2 l_3} 
&\equiv& \sqrt{\frac{(2 l_1 + 1)(2 l_2 + 1)(2 l_3 + 1)}{4 \pi}}
\left(
  \begin{array}{ccc}
  l_1 & l_2 & l_3 \\
   s_1 & s_2 & s_3 
  \end{array}
 \right) ~.
\end{eqnarray}
The resulting $I$ symbol, namely $I_{\ell_1 \ell_2 L}^{0~ 0~ 0}$ (see also eq.~\eqref{eq:Cl_SS}), vanishes except for $\ell_1 + \ell_2 + L = {\rm even}$ and $|\ell_1 - L| \leq \ell_2 \leq \ell_1 + L$, for the selection rules of the $3j$ symbol. This fact and the restriction of $L = 0$ or $2$ in the curvature power spectrum~\eqref{eq:zeta_zeta_1_sYlm} impose the parity-even configurations of CMB power spectrum: $\ell_1 = \ell_2 \pm 2, \ell_2$. This is a consequence of the fact that the scalar mode has no spin dependence and hence cannot create parity asymmetry. On the other hand the off-diagonal correlations arise specifically from breaking statistical isotropy. 

Figure~\ref{fig:Cl_II} describes these power spectra satisfying $m_1 = m_2 = 1$ and $L = 2$, together with the tensor-tensor and scalar-tensor power spectra discussed below. From this figure, one can confirm highly red-tilted spectra ($\ell_1^2 C_{\ell_1 m_1 \ell_2 m_2} \propto \ell_1^{-4}$) satisfying $\ell_1 = \ell_2 \pm 2, \ell_2$.

\subsection{Tensor-tensor correlation}

The tensor temperature fluctuation is expressed with the sum of the two spin states: \cite{Shiraishi:2010sm, Shiraishi:2010kd}
\begin{eqnarray} 
a_{T,\ell m}^{(t)} = 
4\pi (-i)^{\ell} \int \frac{d^3 {\bf k}}{(2\pi)^{3/2}}
 {\cal T}_{T,\ell}^{(t)}(k) \sum_{\lambda = \pm 2} 
 h_{\bf k}^{(\lambda)} 
{}_{-\lambda}Y_{\ell m}^*(\hat{\bf k})~, \label{eq:alm_T}
\end{eqnarray} 
where ${\cal T}_{T,\ell}^{(t)}(k)$ denotes the radiation transfer function of tensor temperature modes. The 0-mode spectrum is of the usual isotropic form:
\begin{eqnarray}
\Braket{a_{T,\ell_1 m_1}^{(t)} a_{T,\ell_2 m_2}^{(t) *}}_0 
= 32\pi \epsilon {\cal P}
\int_0^\infty \frac{dk}{k}
\left[ {\cal T}_{T,\ell_1}^{(t)}(k) \right]^2
 \delta_{\ell_1, \ell_2} \delta_{m_1, m_2} ~.
\end{eqnarray}

Like the scalar case, using eq.~\eqref{eq:h_h_1_sYlm}, the 1-mode spectrum can be written as
\begin{eqnarray}
\Braket{a_{T,\ell_1 m_1}^{(t)} a_{T,\ell_2 m_2}^{(t) *}}_1
&=& 
i^{\ell_2 -\ell_1}
4\pi  {\cal P}
\int_0^\infty  \frac{dk}{k}
{\cal T}_{T,\ell_1}^{(t)}(k)  
{\cal T}_{T,\ell_2}^{(t)}(k) \left(\frac{k_0}{k}\right)^4
\nonumber \\ 
&&\times 
\sum_{L = 0,2} \sum_{M}  g_{LM}^{tt}
\int d^2 \hat{\bf k} {}_{-2}Y_{\ell_1 m_1}^*(\hat{\bf k})
{}_{-2}Y_{\ell_2 m_2}(\hat{\bf k}) Y_{LM} (\hat{\bf k}) 
\nonumber \\
&\equiv& \sum_{L=0,2} C_{\ell_1 m_1 \ell_2 m_2}^{tt;TT}(L) ~.
\end{eqnarray}
Here, the angular integral is given by the spin-$(-2)$ spherical harmonics, corresponding to the $\lambda = +2$ polarized nature of the 1-mode gravitational wave power spectrum. Interestingly, a spin-dependent $I$ symbol arising from the angular integral, namely $I_{\ell_1 \ell_2 2}^{-2 2 0}$ (see eq.~\eqref{eq:gaunt} and eq.~\eqref{eq:Cl_TT}), does not vanish even in the $\ell_1 + \ell_2 ={\rm odd}$ configurations, unlike the scalar case; thus we have both parity-even ($\ell_1 = \ell_2 \pm 2, \ell_2$) and parity-odd ($\ell_1 = \ell_2 \pm 1$) signals. The highly red-tilted power spectra with nonzero parity-odd components are clear in figure~\ref{fig:Cl_II}. These power spectra are suppressed by $\epsilon^2$ compared with the scalar-scalar counterparts.

\subsection{Scalar-tensor correlation}

The CMB scalar-tensor correlation only arises from the 1-mode initial power spectrum \eqref{eq:zeta_h_1_sYlm}, reading
\begin{eqnarray}
\Braket{a_{T,\ell_1 m_1}^{(s)} a_{T,\ell_2 m_2}^{(t) *}}_1
&=& i^{\ell_2 -\ell_1}
4\pi  {\cal P}
\int_0^\infty  \frac{dk}{k}
{\cal T}_{T,\ell_1}^{(s)}(k)  
{\cal T}_{T,\ell_2}^{(t)}(k) \left(\frac{k_0}{k}\right)^4 
\nonumber \\ 
&&\times 
\sum_{M} g_{2M}^{st} 
\int d^2 \hat{\bf k} Y_{\ell_1 m_1}^*(\hat{\bf k})
{}_{-2}Y_{\ell_2 m_2}(\hat{\bf k}) 
{}_{2}Y_{2M} (\hat{\bf k}) \nonumber \\
&\equiv& C_{\ell_1 m_1 \ell_2 m_2}^{st;TT}(L=2) 
\end{eqnarray}
Here, the angular integral results in a spin-dependent $I$ symbol: $I_{\ell_1 \ell_2 2}^{0 2 -2}$ from eq.~\eqref{eq:gaunt} (see also eq.~\eqref{eq:Cl_ST}), and this also creates non-vanishing CMB signals on the parity-even ($\ell_1 = \ell_2 \pm 2, \ell_2$) and parity-odd ($\ell_1 = \ell_2 \pm 1$) domains as seen in figure~\ref{fig:Cl_II}. We can also confirm there that these power spectra are suppressed by $\epsilon$ compared with the scalar-scalar counterparts.

\subsection{Sensitivity to $\rho_E^{\rm CMB}$ and $\hat{\bf E}^{\rm in}$}

Using the Fisher matrix generated from the power spectra computed in the previous subsections, we now evaluate  the expected error bars on the $g_{LM}^{ss}$ parameters.
\footnote{See refs.~\cite{Ma:2011ii, Shiraishi:2014owa} for other studies that perform a Fisher matrix analysis of weakly red-tilted ($k^{-1}$ or $k^{-2}$) quadrupolar modulations compared to our $k^{-4}$ case.} 
Here, we take into account the temperature ($X=T$) and E-mode ($X=E$) signals coming from the scalar auto correlation, since the other contributions are suppressed by the slow-roll parameter ($\epsilon < 10^{-2}$ \cite{Ade:2013uln, Ade:2014xna}) and therefore negligible (see eq.~\eqref{eq:consistency}). 
Supposing the 1-mode power spectra are a small correction to the 0-mode ones, an optimal estimator for $g_{LM}^{ss}$ is expressed as \cite{Hanson:2009gu, Hanson:2010gu,Ma:2011ii}
\begin{eqnarray}
\hat{g}^{ss}_{LM} &=& \frac{1}{2} \sum_{L'M'} {\cal F}_{LM,L'M'}^{-1}
\sum_{\ell_1 m_1 \ell_2 m_2} \sum_{X_1 X_2} \bar{a}_{X_1,\ell_1 m_1}^{*}
\frac{\partial C_{\ell_1 m_1 \ell_2 m_2}^{ss;X_1 X_2}}{\partial g_{L'M'}^{ss *}} 
 \bar{a}_{X_2,\ell_2 m_2} ~,
\end{eqnarray} 
where $\bar{a}_{X,\ell m} \equiv \sum_{X'} (C^{-1})_\ell^{XX'} a_{X',\ell m}$ are the CMB coefficients after weighting with the inverse of the (isotropic) covariance matrix.  The Fisher matrix can be diagonalized as 
\begin{eqnarray}
{\cal F}_{LM, L' M'} 
= \frac{1}{2} \frac{\delta_{L, L'} \delta_{M, M'}}{2L+1} \sum_{\ell_1 \ell_2 } \left( I_{\ell_1 \ell_2 L}^{0~0~0} \right)^2
\sum_{ \substack{X_1 X_2 \\  X_1' X_2'} } 
G_{\ell_1 \ell_2}^{X_1 X_2} 
(C^{-1})_{\ell_1}^{X_1 X_1'} (C^{-1})_{\ell_2}^{X_2 X_2'} 
G_{\ell_1 \ell_2}^{X_1' X_2'}~, \label{eq:fish}
\end{eqnarray}
where 
\begin{eqnarray}
G_{\ell_1 \ell_2}^{X_1 X_2} 
&\equiv& 
4\pi  {\cal P}
\int_0^\infty \frac{dk}{k}
{\cal T}_{X_1,\ell_1}^{(s)}(k)  
{\cal T}_{X_2,\ell_2}^{(s)}(k) 
\left( \frac{k_0}{k} \right)^4
 ~, 
\end{eqnarray}
with ${\cal T}_{X,\ell}^{(s)}(k)$ denoting the radiation transfer functions of the scalar mode. Expected $1\sigma$ errors from the observations are given by $\delta g_{LM}^{ss} = {\cal F}_{LM, LM}^{-1/2}$. Before moving to our $k^{-4}$ case, we computed $\delta g_{2M}$ in the more mildly red-tilted cases ($k^0$, $k^{-1}$ and $k^{-2}$) by use of eq.~\eqref{eq:fish}, and we checked the consistency of our findings with previous results \cite{Ma:2011ii}, showing the validity of our numerical computations.

The numerical results in our $k^{-4}$ case are summarized in table~\ref{tab:error}. In the computations, we have ignored any instrumental features, i.e., beam, noise and partial sky mask, for simplicity. However, the error bars are determined by CMB temperature anisotropies and polarization data on very large-scales ($\ell \lesssim 10$) since the spectrum $G_{\ell_1 \ell_2}^{X_1 X_2}$ is highly red-tilted, thus a realistic analysis with full-sky data provided by e.g., WMAP or {\it Planck} is expected to give similar values, with the expected $1 \sigma$ errors determined mainly by cosmic variance. As we can see from table~\ref{tab:error}, in our $k^{-4}$ case, the E-mode polarization does not help to reduce error bars so much due to the smallness of the polarization large-scale signals.
\footnote{In the mildly red-tilted ($k^{-2}$) case, the E-mode contribution can be comparable to the temperature one \cite{Ma:2011ii}.}

From eq.~\eqref{eq:gLM_def} the $1 \sigma$ errors $\delta g_{00}^{ss} = 16$ and $\delta g_{2M}^{ss} = 30$, can be straightforwardly translated into the error bars on the ratio of the energy density of the gauge field when CMB scales cross the horizon during inflation to the pseudoscalar energy density, reading
\begin{eqnarray} 
\delta \left( \frac{\rho_E^{\rm CMB}}{ \rho_\phi} \right) 
&=& 1.7 \times 10^{-5}  
\left(\frac{\xi}{\xi_{\rm max}}\right)^{7} 
e^{-2 \pi (\xi - \xi_{\rm max})}
\left(\frac{\epsilon}{0.01}\right)
  ~, \\  
\delta \left( \frac{\rho_E^{\rm CMB}}{\rho_\phi} Y_{2M}^* (\hat{\bf E}^{\rm in}) \right) 
&=& 4.4 \times 10^{-5} \left(\frac{\xi}{\xi_{\rm max}}\right)^{7} 
e^{-2 \pi (\xi - \xi_{\rm max})}  \left(\frac{\epsilon}{0.01}\right) 
~,
\end{eqnarray}
where $\xi_{\rm max} = \frac{7}{2\pi} \approx 1.1$ denotes $\xi$ maximizing $\xi^7 e^{- 2\pi \xi}$, and we have used $\rho_\phi \simeq 24\pi^2 \epsilon {\cal P} M_p^4 $. One can see from these expressions that when  
e.g., $\epsilon = 0.01$, $\rho_E^{\rm CMB} /  \rho_\phi = 3.4 \times 10^{-5}$ can be measured at 95\%CL. 
In such cases, the backreaction of the gauge field is totally negligible during inflation; namely eq.~\eqref{eq:potential} holds, and thus our estimations ignoring anisotropic effects on the background metric will be approximately valid 
(although small corrections might arise in the primordial power spectra due to the detailed analysis of nontrivial anisotropic effects
\footnote{Private communication from Atsushi Naruko.}).

\begin{table}[t]
\begin{center}
  \begin{tabular}{|c||c|c|c|} \hline
    & TT & EE & TT+TE+EE \\ \hline
    $\delta g_{00}^{ss}$ & 18 & 31 & 16 ($2.6 \times 10^{-5}$) \\ 
    $\delta g_{2M}^{ss}$ & 33 & 59 & 30 ($4.9 \times 10^{-5}$) \\ \hline
  \end{tabular}
\end{center}
\caption{Expected $1\sigma$ errors on $g_{00}^{ss}$ and $g_{2M}^{ss}$ estimated in the Fisher matrix analysis with the temperature power spectrum (TT), the E-mode power spectrum (EE), and their possible combinations (TT+TE+EE). Here we do not take into account any experimental uncertainties being noticeable on small scales, since $\delta g_{00}^{ss}$ and $\delta g_{2M}^{ss}$ are mostly determined by the cosmic variance on very large scales ($\ell \lesssim 10$). These values can be straightforwardly translated into $\delta g_{LM}$ analyzed with another pivot scale $k_*$ by following $\delta g_{LM} = \delta g_{LM}^{ss}  (k_* \times 14~{\rm Gpc})^{-4}$. For reference, in the brackets, we also describe the corresponding values based on the WMAP or {\it Planck} pivot scale: $k_* = 0.002~{\rm Mpc^{-1}}$ \cite{Hinshaw:2012aka, Ade:2013zuv, Ade:2013uln}.}\label{tab:error}
\end{table}

\section{Summary and discussion} \label{sec:conclusion}

Inflationary models involving a pseudoscalar field yield observationally interesting signatures related to cosmic parity violation. To date, relevant phenomenological analyses have been done under the assumption of isotropy of the Universe. In this work we presented the first analysis of the cosmological signatures taking into account small violation of isotropy due to a nonzero vev of a gauge field coupled to the pseudoscalar inflaton field. 

We found that, together with ordinary isotropic terms, auto correlations of primordial scalar and tensor perturbations have extra quadrupolar modulations depending on the direction of the vev of the gauge field. Such quadrupolar anisotropy also induces non-vanishing cross correlations between scalar and tensor modes. Furthermore, the scalar-tensor and tensor-tensor correlations also break parity symmetry, owing to chiral gravitational waves sourced by the gauge field.  

The most phenomenologically interesting result we found is that, the scalar-tensor and tensor-tensor correlations create both parity-even and also parity-odd signals in the CMB off-diagonal domains. Especially, the signals for $\ell_1 = \ell_2 \pm 1$ in TT, TE, EE or BB, and $\ell_1 = \ell_2 \pm 2$ in TB or EB are impossible to be realized if one of parity and isotropy is preserved (see table~\ref{tab:correspond}). In the case of the pseudoscalar inflation we analyzed, although such special signals are relatively weaker than the scalar-scalar contributions due to slow-roll suppressions of gravitational waves, they will become informative observables to assess early Universe models violating both parity and rotational invariances like our case.

In order to simplify our phenomenological analyses, in this work, we adopted a special background solution of the gauge field, such as ${\bf E} \propto a^{-2}$ and ${\bf B} = 0$. This is due to the assumption of the standard $F^2$ kinetic term for the gauge field. One could imagine combining the $\phi F \tilde{F}$ interaction studied here with a modified kinetic term $f(\phi) F^2$, such that the coherent vev does not rapidly decrease. This will lead to a different scaling of the anisotropic correlators, and to different bounds. We found that in our set-up, the off-diagonal signatures may be observable for energy in the vector field as small as $\sim 10^{-5} \rho_\phi$. It would be indeed interesting to study the phenomenology with a different kinetic function, or in other models in which the background asymmetry can be more relevant.

\acknowledgments
We thank Atsushi Naruko for useful discussions. MS was supported in part by a Grant-in-Aid for JSPS Research under Grant No.~25-573. This work was supported in part by the ASI/INAF Agreement I/072/09/0 for the Planck LFI Activity of Phase E2. The work of MP was supported in part by DOE grant de-sc0011842 at the University of Minnesota.

\appendix

\section{Full temperature and polarization power spectra}\label{appen:CMB_pol}

Here, let us discuss the CMB 1-mode power spectra involving polarizations. The temperature and polarization coefficients of scalar and tensor modes are given by \cite{Shiraishi:2010sm, Shiraishi:2010kd}
\begin{eqnarray}
a_{T/E, \ell m}^{(s)} &=& 
4\pi (-i)^{\ell} \int \frac{d^3 {\bf k}}{(2\pi)^{3/2}}
 {\cal T}_{T/E, \ell}^{(s)}(k) 
 \zeta_{\bf k} Y_{\ell m}^*(\hat{\bf k})~, \\
a_{T/E, \ell m}^{(t)} &=& 
4\pi (-i)^{\ell} \int \frac{d^3 {\bf k}}{(2\pi)^{3/2}}
 {\cal T}_{T/E, \ell}^{(t)}(k)
\left[ h_{\bf k}^{(+2)} {}_{-2}Y_{\ell m}^*(\hat{\bf k}) 
+ h_{\bf k}^{(-2)} {}_{2}Y_{\ell m}^*(\hat{\bf k}) \right]~, \\
a_{B, \ell m}^{(t)} &=& 
4\pi (-i)^{\ell} \int \frac{d^3 {\bf k}}{(2\pi)^{3/2}}
 {\cal T}_{B, \ell}^{(t)}(k)
\left[ h_{\bf k}^{(+2)} {}_{-2}Y_{\ell m}^*(\hat{\bf k}) 
- h_{\bf k}^{(-2)} {}_{2}Y_{\ell m}^*(\hat{\bf k}) \right]~,
\end{eqnarray}
where ${\cal T}_{T/E/B, \ell}^{(s/t)}(k)$ is the radiation transfer function of each mode.
In the same manner as the temperature case, using these $a_{\ell m}$, all types of angular power spectra are formulated as 
\begin{eqnarray}
\Braket{a_{X_1, \ell_1 m_1}^{(s)} a_{X_2, \ell_2 m_2}^{(s) *}}_1
&=& i^{\ell_2 -\ell_1}
4\pi  {\cal P}
\int_0^\infty \frac{dk}{k}
{\cal T}_{X_1,\ell_1}^{(s)}(k)  
{\cal T}_{X_2,\ell_2}^{(s)}(k) \left( \frac{k_0}{k} \right)^4 
\nonumber \\ 
&& \times 
\sum_{L=0,2} \sum_{M} g_{LM}^{ss}
(-1)^{m_1}
I_{\ell_1 \ell_2 L}^{0~0~0}
 \left(
  \begin{array}{ccc}
  \ell_1 & \ell_2 & L \\
  -m_1 & m_2 & M
  \end{array}
 \right) \nonumber \\
&\equiv& \sum_{L=0,2} C_{\ell_1 m_1 \ell_2 m_2}^{ss; X_1 X_2}(L) ~, \label{eq:Cl_SS} \\
\Braket{a_{X_1, \ell_1 m_1}^{(t)} a_{X_2, \ell_2 m_2}^{(t) *}}_1 
&=& i^{\ell_2 -\ell_1}
4\pi  {\cal P}
\int_0^\infty \frac{dk}{k}
{\cal T}_{X_1,\ell_1}^{(t)}(k)  
{\cal T}_{X_2,\ell_2}^{(t)}(k) \left( \frac{k_0}{k} \right)^4 
\nonumber \\ 
&& \times 
\sum_{L=0,2} \sum_{M} g_{LM}^{tt}
(-1)^{m_1}
I_{\ell_1 \ell_2 L}^{-2 2 0}
 \left(
  \begin{array}{ccc}
  \ell_1 & \ell_2 & L \\
  -m_1 & m_2 & M
  \end{array}
 \right) \nonumber \\
&\equiv& \sum_{L=0,2} C_{\ell_1 m_1 \ell_2 m_2}^{tt; X_1 X_2}(L) 
~, \label{eq:Cl_TT} \\
\Braket{a_{X_1, \ell_1 m_1}^{(s)} a_{X_2, \ell_2 m_2}^{(t) *}}_1
&=& i^{\ell_2 -\ell_1}
4\pi  {\cal P}
\int_0^\infty \frac{dk}{k}
{\cal T}_{X_1,\ell_1}^{(s)}(k)  
{\cal T}_{X_2,\ell_2}^{(t)}(k) \left( \frac{k_0}{k} \right)^4 
\nonumber \\ 
&& \times 
\sum_{M} g_{2M}^{st}
(-1)^{m_1}
I_{\ell_1 \ell_2  2}^{0 2-2}
 \left(
  \begin{array}{ccc}
  \ell_1 & \ell_2 & 2 \\
  -m_1 & m_2 & M
  \end{array}
 \right) \nonumber \\ 
&\equiv& C_{\ell_1 m_1 \ell_2 m_2}^{st; X_1 X_2}(L = 2) 
~. \label{eq:Cl_ST}
\end{eqnarray}

\begin{figure}[t]
  \begin{tabular}{c}
    \begin{minipage}{1.0\hsize}
  \begin{center}
    \includegraphics[width = 0.5\textwidth]{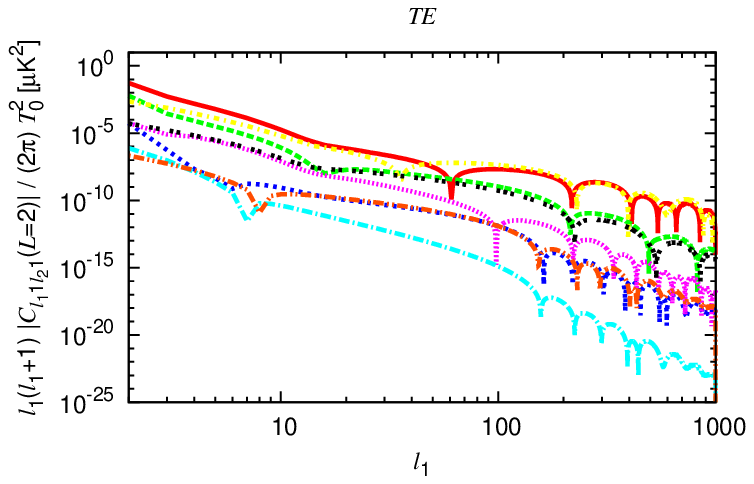}
  \end{center}
\end{minipage}
\end{tabular}
\\
  \begin{tabular}{cc}
    \begin{minipage}{0.5\hsize}
  \begin{center}
    \includegraphics[width=1\textwidth]{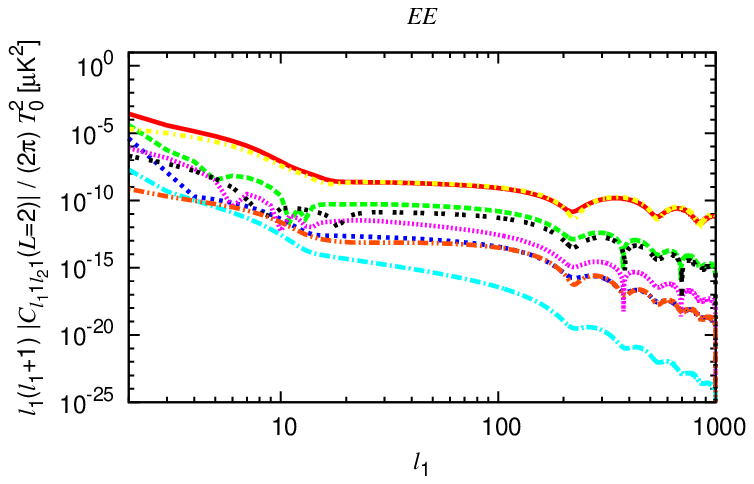}
  \end{center}
\end{minipage}
\begin{minipage}{0.5\hsize}
  \begin{center}
    \includegraphics[width=1\textwidth]{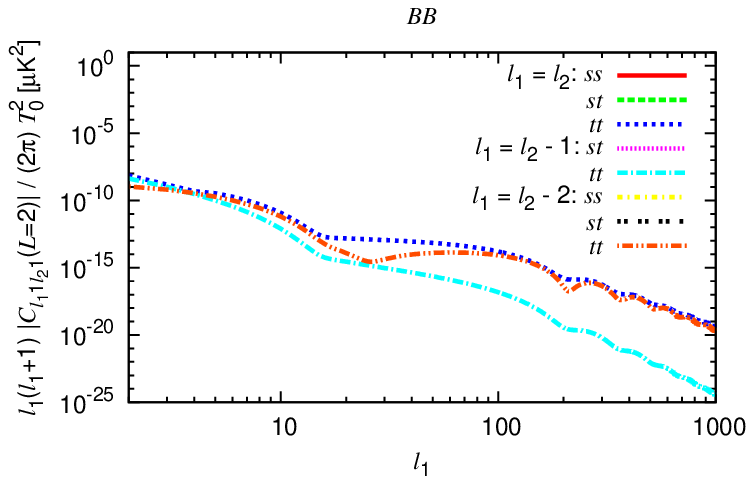}
  \end{center}
\end{minipage}
\end{tabular}
\\
  \begin{tabular}{cc}
    \begin{minipage}{0.5\hsize}
  \begin{center}
    \includegraphics[width=1\textwidth]{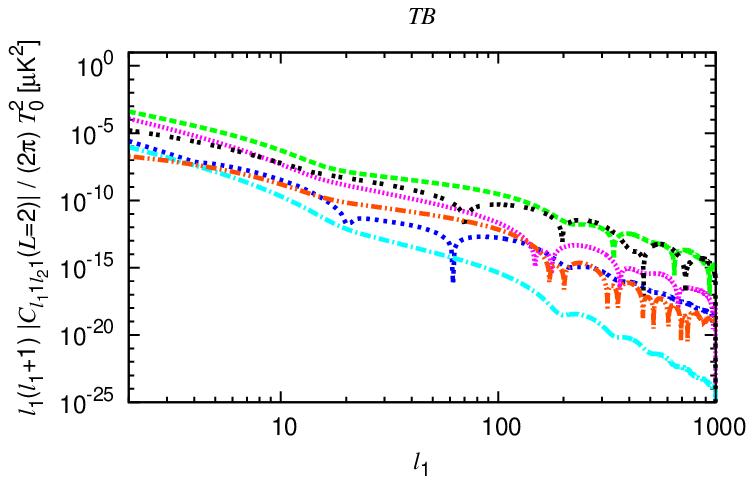}
  \end{center}
\end{minipage}
\begin{minipage}{0.5\hsize}
  \begin{center}
    \includegraphics[width=1\textwidth]{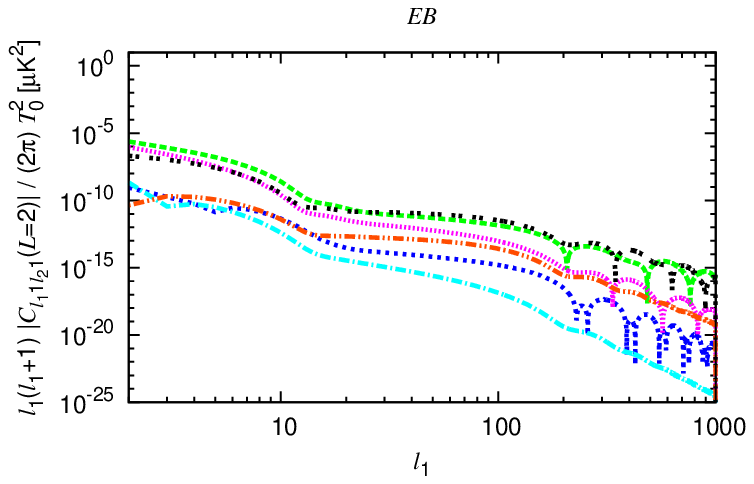}
  \end{center}
\end{minipage}
\end{tabular}
  \caption{All possible 1-mode power spectra including polarizations: $C_{\ell_1 1 \ell_2 1}^{z_1 z_2; X_1 X_2} (L = 2)$. The settings are identical to figure~\ref{fig:Cl_II}.} \label{fig:Cl_pol}
\end{figure}

Figures~\ref{fig:Cl_II} and \ref{fig:Cl_pol} describe all possible correlations coming from the quadrupolar ($L=2$) coefficient. It is verified from these figures that the highly red-tilted spectra appear in both $\ell_1 = \ell_2 \pm 2, \ell_2$ and $\ell_1 = \ell_2 \pm 1$, even for the polarization cases. 
The non-vanishing special signals associated with broken parity and rotational invariance, namely $\ell_1 = \ell_2 \pm 1$ in TE, EE and BB, and $\ell_1 = \ell_2 \pm 2$ in TB and EB, are allowed by the selection rules of the $I$ symbols in the similar manner as the temperature case (see section~\ref{sec:CMB}). 
Note that the spectrum including B mode, i.e, TB/EB or BB, is sourced from one or two tensor mode, and is suppressed by $\epsilon$ or $\epsilon^2$. This fact decreases the signal-to-noise ratios in comparison with the temperature and E-mode analyses.

\section{Polarization vector and tensor}\label{appen:pol}

The divergenceless polarization vector ($\epsilon_a^{(\pm 1)}$) and transverse and traceless polarization tensor ($e_{ab}^{(\pm 2)}$) adopted in this paper obey the following conditions \cite{Shiraishi:2010kd}:
\begin{eqnarray}
\begin{split}
\hat{k}_a \epsilon_a^{(\lambda)}(\hat{\bf k}) &= 0~, \\ 
\eta^{0abc} \hat{k}_a \epsilon_b^{(\lambda)}(\hat{\bf k}) 
&= -\lambda i \epsilon_c^{(\lambda)}(\hat{\bf k}) ~, \\ 
\epsilon^{(\lambda) *}_a (\hat{\bf k}) &= \epsilon^{(-\lambda)}_a (\hat{\bf k})
 = \epsilon^{(\lambda)}_a (-\hat{\bf k})~, \\
\epsilon^{(\lambda)}_a (\hat{\bf k}) \epsilon^{(\lambda')}_a (\hat{\bf k}) 
&= \delta_{\lambda, -\lambda'} ~, 
\end{split}
\end{eqnarray}
and 
\begin{eqnarray}
\begin{split}
e_{ab}^{(\lambda)}(\hat{\bf k}) &\equiv \sqrt{2} \epsilon_a^{(\frac{\lambda}{2})}(\hat{\bf k}) \epsilon_b^{(\frac{\lambda}{2})}(\hat{\bf k}) ~, \\ 
e_{aa}^{(\lambda)}(\hat{\bf k}) &= \hat{k}_a e_{ab}^{(\lambda)}(\hat{\bf k}) = 0~, \\
e_{ab}^{(\lambda) *}(\hat{\bf k}) &= e_{ab}^{(-\lambda)}(\hat{\bf k}) = e_{ab}^{(\lambda)}(- \hat{\bf k})~, \\
e_{ab}^{(\lambda)}(\hat{\bf k}) e_{ab}^{(\lambda')}(\hat{\bf k}) &= 2
\delta_{\lambda, -\lambda'}~. \label{eq:pol_tens_relation}
\end{split}
\end{eqnarray}
Harmonic representations are convenient to contract between unit vectors, polarization vectors, and polarization tensors, reading \cite{Shiraishi:2010kd}
\begin{eqnarray}
\hat{k}_a &=& \sum_m \alpha_a^{m} Y_{1 m}(\hat{\bf k}) ~, \\ 
\epsilon_a^{(\lambda)} (\hat{\bf k}) 
&=& -\lambda \sum_m \alpha_a^m {}_{\lambda} Y_{1 m} (\hat{\bf k})  ~, \\
e_{ab}^{(\lambda)} (\hat{\bf k}) 
&=& \frac{3}{\sqrt{2 \pi}}  
\sum_{M m_a m_b} {}_{-\lambda}Y_{2 M}^*(\hat{\bf k}) 
\alpha^{m_a}_{a} \alpha^{m_b}_b 
\left(
  \begin{array}{ccc}
  2 & 1 &  1\\
  M & m_a & m_b 
  \end{array}
\right) ~,
\end{eqnarray}
with
\begin{eqnarray}
\alpha_a^m \alpha_a^{m'} = \frac{4 \pi}{3} (-1)^m \delta_{m,-m'}~, \ \
\alpha_a^m \alpha_a^{m' *} = \frac{4 \pi}{3} \delta_{m,m'}~.
\end{eqnarray}
Using these relations, we can easily derive the harmonic representations adopted in eqs.~\eqref{eq:zeta_zeta_1_sYlm}, \eqref{eq:h_h_1_sYlm} and \eqref{eq:zeta_h_1_sYlm}:
\begin{eqnarray}
 \sum_{s= \pm 1} 
\epsilon_i^{(s)}(\hat{\bf k}) \epsilon_j^{(-s)}(\hat{\bf k})
 \hat{E}_i^{\rm in} \hat{E}_j^{\rm in} 
&=& 1 - \left(\hat{\bf k} \cdot \hat{\bf E}^{\rm in} \right)^2  \nonumber \\
&=& 
\frac{8\pi}{3} \sum_{L=0,2} \sum_{M} \frac{i^L}{2L+1}
Y_{LM}^* (\hat{\bf E}^{\rm in}) 
Y_{LM} (\hat{\bf k})
~, \\
e_{ij}^{(\lambda)}(\hat{\bf k}) \hat{E}_i^{\rm in} \hat{E}_j^{\rm in} 
&=& 
 \frac{8\sqrt{3} \pi}{15}  \sum_{M}
 Y_{2 M}^*(\hat{\bf E}^{\rm in}) {}_{\lambda}Y_{2 M}(\hat{\bf k}) \label{eq:identity_e2EE}
~.
\end{eqnarray}

\bibliography{paper}
\end{document}